\documentclass[prx, preprint, superscriptaddress, endfloats*]{revtex4-2}
\usepackage{amsmath,hyperref,cleveref,amssymb,amsfonts,graphicx,multirow,color,bm,textcomp,mathtools}
\usepackage{paralist}
\usepackage{srcltx}
\usepackage[all,cmtip]{xy}
\usepackage{mathrsfs}
\usepackage{srcltx,soul,subfigure}
\usepackage{threeparttable,dsfont,bbm}
\usepackage{color}
\usepackage{comment}
\usepackage{dcolumn}

\begin{document}
\title{Emergent Spin Fluctuation and Structural Metastability in Self-Intercalated Cr$_{1+x}$Te$_2$ Compounds}
\author{Clayton Conner}
\affiliation {Department of Physics and Astronomy, University of Missouri, Columbia, Missouri 65211, USA}
\author{Ali Sarikhani}
\affiliation {Department of Physics and Astronomy, Missouri University of Science and Technology, Rolla, Missouri 65409, USA}
\author{Theo Volz}
\affiliation {Rock Bridge High School, Columbia, Missouri 65203, USA}
\author{Mathew Pollard}
\affiliation {Department of Physics and Astronomy, Missouri University of Science and Technology, Rolla, Missouri 65409, USA}
\author{Mitchel Vaninger}
\affiliation {Department of Physics and Astronomy, University of Missouri, Columbia, Missouri 65211, USA}
\author{Xiaoqing He}
\affiliation {Electron Microscopy Core Facility, University of Missouri, Columbia, Missouri 65211, USA}
\author{Steven Kelley}
\affiliation {Department of Chemistry, University of Missouri, Columbia, Missouri 65211, USA}
\author{Jacob Cook}
\affiliation {Department of Physics and Astronomy, University of Missouri, Columbia, Missouri 65211, USA}
\author{Avinash Sah}
\affiliation {Department of Physics and Astronomy, University of Missouri, Columbia, Missouri 65211, USA}
\author{John Clark}
\affiliation {Department of Physics and Astronomy, University of Missouri, Columbia, Missouri 65211, USA}
\author{Hunter Lucker}
\affiliation {Department of Physics and Astronomy, University of Missouri, Columbia, Missouri 65211, USA}
\author{Cheng Zhang}
\affiliation {Department of Physics and Astronomy, University of Missouri, Columbia, Missouri 65211, USA}
\author{Paul Miceli}
\affiliation {Department of Physics and Astronomy, University of Missouri, Columbia, Missouri 65211, USA}
\author{Yew San Hor}
\affiliation {Department of Physics and Astronomy, Missouri University of Science and Technology, Rolla, Missouri 65409, USA}
\author{Xiaoqian Zhang}
\affiliation {Key Laboratory of Quantum Materials and Devices of Ministry of Education, School of Physics, Southeast University, Nanjing 211189, China}
\author{Guang~Bian}\email{biang@missouri.edu}
\affiliation {Department of Physics and Astronomy, University of Missouri, Columbia, Missouri 65211, USA}
\affiliation {MU Materials Science \& Engineering Institute, University of Missouri, Columbia, MO, 65211, USA}

\newpage

\begin{abstract}
Intercalated van der Waals (vdW) magnetic materials host unique magnetic properties due to the interplay of competing interlayer and intralayer exchange couplings, which depend on the intercalant concentration within the van der Waals gaps. Magnetic vdW compound chromium telluride, Cr$_{1+x}$Te$_2$, has demonstrated rich magnetic phases at various Cr concentrations, such as the coexistence of ferromagnetic and antiferromagnetic phases in Cr$_{1.25}$Te$_2$ (equivalently, Cr$_{5}$Te$_8$). The compound is induced by intercalating 0.25 Cr atom per unit cell within the van der Waals gaps of CrTe$_2$. In this work, we report a notably increased Curie Temperature and an emergent in-plane spin fluctuation by slightly reducing the concentration of Cr intercalants in Cr$_{1.25}$Te$_2$. Moreover,  the intercalated Cr atoms form a metastable 2$\times$2 supercell structure that can be manipulated by electron beam irradiation. This work offers a promising approach to tuning magnetic and structural properties by adjusting the concentration of intercalated magnetic atoms.

\end{abstract}

\pacs{}%

\maketitle

\newpage

\section{Introduction}

2D magnetic materials have become an intense focus of research in the field of condensed matter physics due to their unique magnetic, electronic, and optical properties. Transition-metal dichalcogenides (TMDs) are layered structures with van der Waals (vdW) gaps between subsequent atomic layers. The two-dimensional layered vdW structure provides the ability to produce atomically thin films with various desired characteristics \cite{CHOI2017116}. VdW magnetic materials hold great potential for novel spintronic applications owing to their relative ease of production, thickness-dependent magnetic properties, and persistence of magnetic ordering down to the two-dimensional regime \cite{KHAN2020100902,Burch2018-tl}. Materials such as Cr$_2$Ge$_2$Te$_6$ \cite{Gong2017-iw}, Fe$_3$GeTe$_2$ \cite{PhysRevB.93.134407,PhysRevB.93.144404}, and CrI$_3$ \cite{Huang2017-vl} have all been heavily studied and show magnetic ordering that persists in the two-dimensional limit; however, many of these materials have a Curie Temperature ($T_\mathrm{C}$) much lower than room temperature, making them largely impractical for device applications. Among the known vdW materials, one of the promising candidates for room-temperature spintronics is the transition-metal dichalcogenide CrTe$_2$, which has been reported to have magnetic ordering down to one monolayer with a $T_\mathrm{C}$ close to room temperature \cite{Zhang2021-cm}.


Magnetic TMDs contain vdW gaps between adjacent layers that can be intercalated with transition metals, creating new compounds that maintain the layered, quasi-2D structure. These compounds can have unique properties that are quite different from the parent vdW material. This intercalation method can be utilized to tune desirable magnetic properties for device applications\cite{Rajapakse2021-fj,Zhao2020-rd}. This tunability through intercalation has been well demonstrated in chromium telluride compounds, denoted as  Cr$_{1+x}$Te$_2$. These compounds have the layered base structure of CrTe$_2$, belonging to the space group P3$\overline{m}$ \cite{Meng2021-wm}, in which Cr atoms can be intercalated within the vdW gaps between the CrTe$_2$ layers, thus creating new compounds \cite{PhysRevMaterials.4.114001}. For example, $x$ = 0 (CrTe$_2$), 0.25 (Cr$_5$Te$_8$), 0.33 (Cr$_2$Te$_3$), 0.50 (Cr$_3$Te$_4$), and 1 (CrTe). Due to the self-intercalation of Cr within the vdW gaps, the properties of these compounds differ significantly from the pristine CrTe$_2$ base compound. As a result, the family of Cr$_{1+x}$Te$_2$ compounds show a variation of $T_\mathrm{C}$, magnetic ordering, and even structural properties \cite{PhysRevB.100.024434,IPSER1983265,JDijkstra_1989}. This wide range of properties necessitates a systematic study of how self-intercalation can be used to manipulate the properties of vdW compounds. 

One of the intercalated variants, Cr$_5$Te$_8$, which is ferromagnetic with a $T_\mathrm{C}$ of approximately 150 K in bulk\cite{LUKOSCHUS2004951}, has been reported to exhibit intriguing magnetic properties such as colossal anomalous Hall effect \cite{Tang2022-bl, doi:10.1021/acsnano.4c08700} and thickness-dependent 
$T_\mathrm{C}$ \cite{https://doi.org/10.1002/adma.202107512}. Furthermore, an emergent antiferromagnetic (AFM) phase with $T_\mathrm{N}\sim 170$~K in Cr$_5$Te$_8$ has been reported, which demonstrates a large negative MR that enables effectively a spin valve device \cite{https://doi.org/10.1002/adfm.202202977, doi:10.1021/acsnano.4c08700}. In this work, we modulate chemical composition of these Cr$_{1+x}$Te$_2$ materials through a self-intercalating approach and explore the structural and magnetic properties of stoichiometries with $x < 0.25$. Our experiments demonstrate that a slightly reduced Cr concentration from Cr$_5$Te$_8$ leads to a metastable structure of intercalants, a significant increase in $T_\mathrm{c}$, and an emergent in-plane spin fluctuation. 

\section{Results and Discussion}

Single crystals of Cr$_{1+x}$Te$_2$ of two different Cr concentrations ($x$ = 0.223 and $x$ = 0.202) were fabricated using the chemical vapor transport (CVT) method. It corresponds to a slight reduction of Cr concentration from Cr$_5$Te$_8$ ($x$ = 0.25), giving the stoichiometries of Cr$_{4.89}$Te$_8$ and Cr$_{4.81}$Te$_8$, respectively. These single-crystal samples serve as a platform for investigating the structural and magnetic properties through varying the concentration of intercalated Cr within the vdW gaps. X-ray diffraction (XRD) measurements reveal the formation of a monoclinic crystal structure belonging to the space group I2/m. These compounds share a similar structure to Cr$_5$Te$_8$ with self-intercalated atoms within the vdW gap of the CrTe$_2$ base layers; however, these compounds contain two inequivalent Cr sites with partial occupancy located within the vdW gap. These Cr sites are labeled as Cr$_4$ and Cr$_7$ in Figure 1(a), with the Cr$_4$ site having a higher statistical occupancy than the Cr$_7$ site (Table 1). In this context, statistical occupancy refers to the probability of finding an intercalated Cr atom located at the respective site. The monoclinic Cr$_5$Te$_8$ crystal structure, on the other hand, contains one fully occupied Cr site located between the CrTe$_2$ layers. These Cr$_4$ and Cr$_7$ intercalated sites create a 2$\times$2 periodicity along in-plane directions due to the inequivalent occupancy of the sites. The lattice constants were determined using single-crystal XRD for both compounds, respectively, as shown in Table 1. We can directly compare these measured lattice constants to the reported monoclinic Cr$_5$Te$_8$ structure of $a$ = 13.575 \AA, $b$ = 7.859 \AA, and $c$ = 12.073 \AA \cite{BENSCH1997305}. While $a$ and $b$ are almost unchanged, $c$ show a reduction of  0.77\%  and 0.93\% for Cr$_{1+x}$Te$_2$  ($x$ = 0.223, 0.202), respectively. This is due to the reduced concentration of Cr intercalants. As the $c$ lattice parameter reflects the coupling between CrTe$_2$ layers, the result suggests an increase in interlayer coupling compared to that of Cr$_5$Te$_8$.

Transmission Electron Microscopy (TEM) and Scanning Transmission Electron Microscopy (STEM) were used to fully characterize the structure of these Cr$_{1+x}$Te$_2$ samples. High-angle annular dark-field (HAADF)-STEM was performed on this sample, and the result is shown in Figure 1(b). The overlaid XRD structure (marked by the yellow and blue balls) aligns well with the HAADF-STEM measurements. Figure 1(c) shows a cross-sectional TEM measurement of  Cr$_{4.89}$Te$_8$ along the (010) direction, confirming the layered quasi 2D structure consistent with the XRD results. An out-of-plane (OOP) TEM image of the lattice can be seen in Figure 1(d), which is in good agreement with the XRD results. The high structural quality of this compound is further confirmed by the selected area electron diffraction (SAED) measurement as shown in Figs.~1(e,f). The corresponding hexagonal periodicity of the SAED result is indexed according to the structure determined by XRD. 

Fast Fourier Transform (FFT) was applied to the cross-sectional TEM image to show the periodicity of the lattice along the (010) direction (Figure 2). This FFT result exhibited the expected periodicity and high degree of crystallinity as indicated by the bright spots (Figure 2a). The FFT image was then indexed according to the XRD crystal structure, as noted in the figure. In addition to the expected periodicity, a second periodicity can be seen as indicated by the dimmer spots (highlighted by the yellow circles). Those spots are located closer to the (000) peak, indicative of a superstructure in real space. This superstructure in reciprocal space has a 2$\times$2 periodicity that matches exactly with the 2$\times$2 periodicity of the intercalated Cr sites, Cr$_4$ and Cr$_7$, as shown in the structure above the FFT image in Figure 2(a). Interestingly, the metastability of this superstructure was observed during the TEM measurements. After exposure to the electron beam from TEM, the 2$\times$2 superstructure spots begin to fade in brightness, as seen in Figure 2(b). After several minutes under electron beam irradiation, the superstructure disappears completely (Figure 2(c)). This effect implies that the Cr$_4$ and Cr$_7$ sites have much lower binding energy than the Cr atoms in the CrTe$_2$ main layer. The scattering with the electron beam provides Cr intercalants enough energy to overcome the binding energy, allowing the Cr intercalants to move randomly within the van der Waals gap and destroy the superstructure periodicity, as schematically depicted in Figs. 2(b,c). Turning off the electron beam, the superstructure is restored after several minutes as shown in Figure 2(d), indicating the metastability of the 2$\times$2 superstructure of Cr intercalants.

The effects of altering the concentration of Cr intercalants were studied by the measurements of magnetic properties of the Cr$_{1+x}$Te$_2$ compounds. The magnetic hysteresis loops at various temperatures (10$-$300~K) for the Cr$_{4.89}$Te$_8$ and Cr$_{4.81}$Te$_8$ compounds can be seen in Figure 3.  The OOP field measurement shows that the magnetization fully saturates at approximately 3000~Oe for both Cr$_{4.89}$Te$_8$ and Cr$_{4.81}$Te$_8$ compounds (Figs.~3(a,c)). By contrast, neither compound fully saturates under an in-plane (IP) magnetic field even up to a field strength of 6 Tesla (Figs.~3(b,d)). This confirms a large perpendicular magnetic anisotropy (PMA), an essential ingredient for stabilizing ferromagnetism in 2D materials \cite{Gibertini2019-sr}, and an out-of-plane magnetic easy axis in both compounds. This behavior can be attributed to the parallel exchange coupling within the CrTe$_2$ layers. It is interesting to note that both Cr$_{4.89}$Te$_8$ and Cr$_{4.81}$Te$_8$ compounds have negligible coercivity, even at very low temperatures, consistent with the typical small coercivity value previously found in Cr$_5$Te$_8$ \cite{Zhang2021-cm, LUKOSCHUS2004951}. 

The temperature-dependent magnetization, $M(T)$ of the two self-intercalated compounds was measured at various magnetic field strengths over a temperature range from 50 K to 300 K.  Figures 4(a,c) show the OOP field measurement of the Cr$_{4.89}$Te$_8$ and Cr$_{4.81}$Te$_8$ compounds, with both compounds exhibiting the characteristic ferromagnetic behavior. The magnetization increases suddenly as the temperature is reduced below the $T_\mathrm{C}$ of 172~K. In addition to the ferromagnetic transition, we observed the appearance of a subtle kink feature at 180 K and 240 K for Cr$_{4.89}$Te$_8$ and Cr$_{4.81}$Te$_8$, respectively. A similar cusp-like feature has been previously reported as an OOP antiferromagnetic ordering in Cr$_5$Te$_8$ \cite{https://doi.org/10.1002/adfm.202202977}. However, the kink feature observed in Cr$_{4.89}$Te$_8$ and Cr$_{4.81}$Te$_8$ is much weaker than that of Cr$_5$Te$_8$ and thus is unlikely to be generated by an emergent AFM phase in the OOP direction. 

The measurements with in-plane fields on the two compounds exhibit a similar FM behavior in the magnetization but with a much lower magnetization value than the OOP case, indicating an OOP easy axis of magnetization. On the other hand, the in-plane $M(T)$ curves exhibit a pronounced cusp-like feature, compared to the OOP results, as shown in Figs.~4(b,d). This cusp feature signals an emergent spin fluctuation in response to the applied IP magnetic field. The spin fluctuation reduces the slope of the M(T) curves as the temperature decreases. This spin fluctuation occurs at $T'=$ 245~K and 229~K for Cr$_{4.89}$Te$_8$ and Cr$_{4.81}$Te$_8$, respectively. The $M(T)$ result indicates that slightly reducing the concentration of Cr intercalants leads to an IP spin fluctuation rather than an OOP AFM ordering. This feature highlights the sensitivity of the exchange couplings of Cr atoms to the chemical composition in this family of magnetic vdW compounds.

One of the most intriguing properties found in the Cr$_5$Te$_8$ compound was the large negative magnetoresistance (MR) that occurred between the $T_\mathrm{C}$ and $T_\mathrm{N}$. This effect can be attributed to the coexistence of the FM and AFM phases. This effect has been demonstrated to create a single-crystal MR spin valve device that can be controlled by weak magnetic fields \cite{https://doi.org/10.1002/adfm.202202977}. A similar modulation of the magnetoresistance can be induced by the emergent spin fluctuation in Cr$_{4.89}$Te$_8$ and Cr$_{4.81}$Te$_8$, arising from the enhanced scattering of electrons with fluctuating spin. The measured the longitudinal resistance with varied OOP magnetic fields for the Cr$_{4.81}$Te$_8$ compound is plotted in Fig.~4(e). It can be seen that the measured MR decreases at larger fields due to the alignment of magnetic moments. Upon zooming in on the MR curves, it's clear that a hump feature appears in the MR curves around the zero field in a temperature window of 170$-$200~K and is fully suppressed above 220~K. This hump in the MR curves corresponds to an enhanced MR induced by the spin fluctuation. The temperature window in which the enhanced MR appears coincides with the temperature range between $T_\mathrm{C}$ (172~K) and $T'$ (229~K) of Cr$_{4.81}$Te$_8$, confirming the emergent spin fluctuation phase.  

The origin of this in-plane spin fluctuation phase can be attributed to several potential mechanisms. One possible mechanism for the in-plane nature of the spin fluctuations is that the in-plane spin component of the intercalated Cr has three equivalent minima of the magnetic ground state. The minima correspond to three energetically preferred in-plane spin orientations \cite{Bigi2025}. Therefore,  the spin of intercalated Cr atoms randomly chooses one of three orientations, giving rise to the spin fluctuation phase. In addition, the structural effect on the magnetic properties of these compounds must be considered.  A trend of decreasing $c$ lattice parameter with the reduction of intercalated Cr has been confirmed by our XRD results (Table~1). This $c$ lattice parameter directly reflects the interlayer distance between the CrTe$_2$ base layers.  This reduction of the interlayer spacing can lead to a stronger hybridization between the 3$d$ orbitals of Cr and the 5$p$ orbitals of Te, thus enhancing the superexchange coupling. The superexchange coupling is highly dependent on the bond distances and angles associated with the Cr-Te-Cr paths to make it either AFM or FM type \cite{PhysRevB.92.214419,doi:10.1021/acs.nanolett.8b03321}.

\section{Conclusion}
In this work, we demonstrate structural and magnetic properties of two self-intercalated Cr$_{1+x}$Te$_2$ compounds with slightly reduced concentration of Cr intercalants compared to that of Cr$_5$Te$_8$ ($x$ = 0.25). A 2$\times$2 superstructure with partially occupied intercalated Cr sites was observed in those compounds. The metastable superstructure can be manipulated by electron beam irradiation in a completely reversible process. Both compounds show an increased Curie temperature ($T_\mathrm{C}$) of 172~K, which is 22~K higher than that of Cr$_5$Te$_8$. Moreover, both compounds exhibit an emergent spin fluctuation above $T_\mathrm{C}$, different from the AFM phase reported in Cr$_5$Te$_8$. In addition, we see a prominent increase in magnetoresistance due to the emergent spin fluctuation phase. This work demonstrates that varying intercalant concentration offers a practical route to tuning the magnetic and structural properties of vdW magnets for spintronic applications. 

\section{Methods}
\subsection{Growth of Cr$_{1+x}$Te$_2$ ($x$ = 0.223, 0.202) compounds}
The Cr$_{1+x}$Te$_2$ samples were produced using a chemical vapor transport (CVT) method, which involves a two-step process to create large single crystals. High-purity chromium (Cr, 99.99\%) powder and tellurium (Te, 99.999\%) lumps were sourced from Alfa Aesar for this growth process. Initially, a stoichiometrically correct mixture of Cr and Te was prepared and placed into a fused quartz ampule, repetitively cleaned with ultra-high-purity (UHP) argon, and sealed under a high vacuum. These ampules were heated to 950$^{\circ}$C for 24 hours, then cooled at a rate of 1$^{\circ}$C per minute to room temperature. The polycrystalline product was then ground into a fine, smooth powder using an agate mortar and pestle, creating a precursor ready for CVT growth.

For the CVT growth, 1 gram of the prepared powdered precursor was placed into a 30 cm-long quartz ampule with a 12 mm inner diameter, along with 3 mg of high-purity iodine pieces per cm$^3$ of internal ampule volume, which serves as the transport agent. The ampules were sealed under the same vacuum and argon conditions as the precursor preparation. To achieve a temperature gradient essential for CVT, a three-zone Thermo Scientific Lindberg Blue M furnace was used. First, zones 2 and 3 of the furnace were preheated to 950°C while zone 1 was maintained at 650$^{\circ}$C for 24 hours to eliminate any impurities on the quartz surface. Following this cleaning phase, the zones were adjusted to 950, 600, and 850$^{\circ}$C respectively, establishing a temperature gradient for efficient vapor transport, with thermocouple readouts recorded as 950, 808, and 850$^{\circ}$C, respectively. This configuration was held steady for two months for a thermal equilibrium as well as larger crystal growth. After the growth phase, the temperature across all zones was gradually equalized to 650$^{\circ}$C and then reduced to room temperature at a rate of 1$^{\circ}$C per minute. Large crystals were cleaved and selected from the bulk for characterization and measurement.

\subsection{Transmission Electron Microscopy and HAADF-STEM}
Lamellar samples were removed from bulk Cr$_{1+x}$Te$_2$ crystals using FIB milling by a FEI Scios DualBeam SEM. TEM and HAADF-STEM images were obtained using a ThermoScientific Spectra 300 electron microscope. 

\subsection{Transport and Magnetic Measurements}
Magnetic characterization was performed using a Quantum Design superconducting quantum interference device (SQUID). This device has a base temperature of 2 K and a magnetic field of up to 7 T. The magnetoresistance measurements were performed on a Quantum Design Physical Property Measurement System (PPMS) with a base temperature of 2 K and magnetic field up to 9 T. The devices were confined to the Hall-bar geometry using a standard four-point probe method with room-temperature cured silver paste contacts.

\section{Acknowledgements}
The work at the University of Missouri was supported by the Gordon and Betty Moore Foundation, grant DOI:10.37807/gbmf12247. 

\bibliographystyle{apsrev4-2}
\bibliography{CrTe.bib}


\begin{figure}
\includegraphics[width=1.0\linewidth]{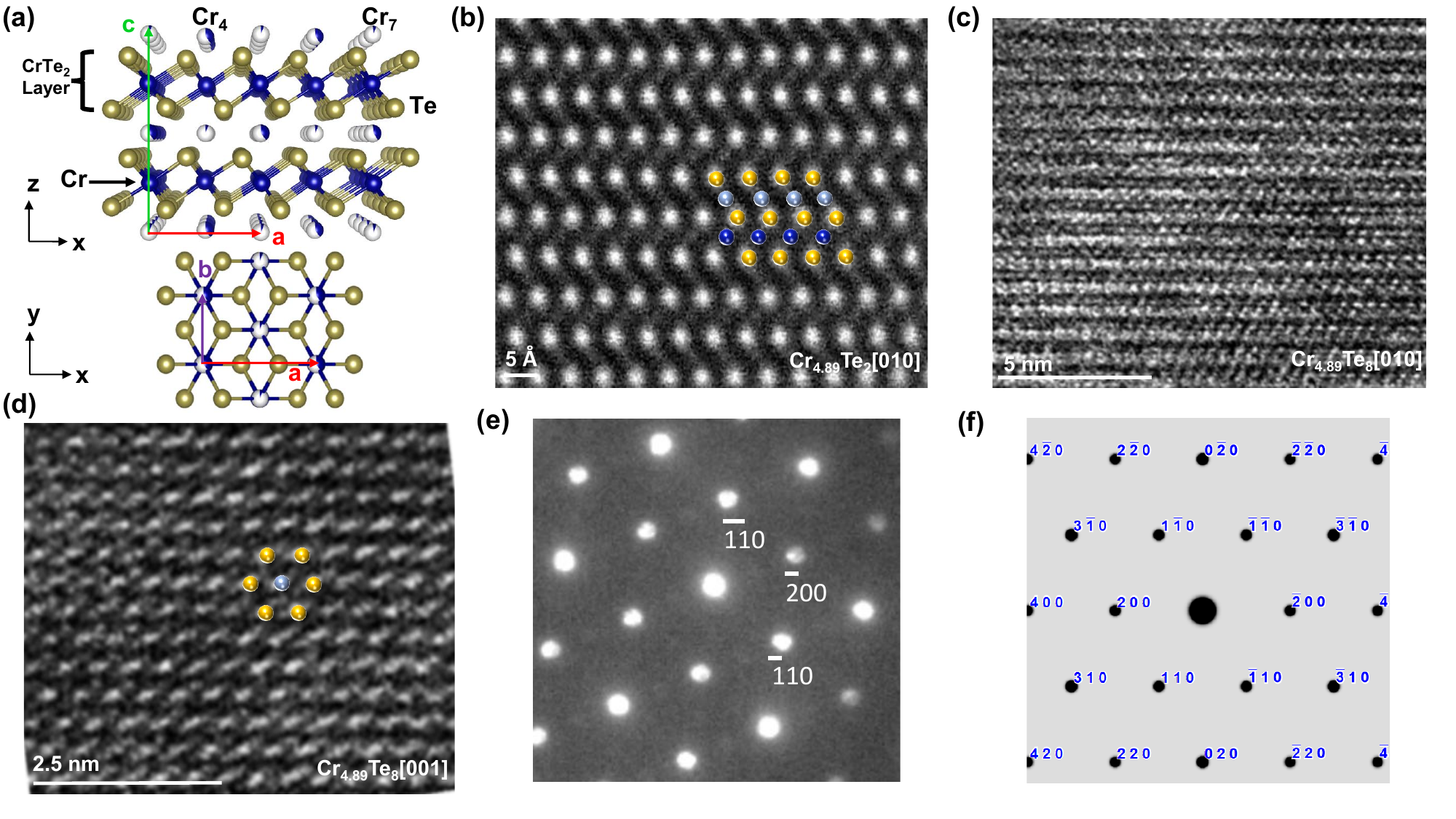}
\caption{(a) Crystal structure of Cr$_{1+x}$Te$_2$ with Cr$_4$ and Cr$_7$ intercalant sites intercalated within vdW gap between CrTe$_2$ layers. A 2$\times$2 periodic superstructure can be seen from these intercalated sites.  (b) HAADF-STEM image of Cr$_{4.89}$Te$_8$ along the (010) direction. (c) Cross-sectional TEM image of Cr$_{4.89}$Te$_8$ along the (010) direction, showing a quasi-2D layered structure that matches with the XRD data shown in (a). (d) The TEM image along the (001) direction that matches the crystal structure shown in (a). (e) Electron diffraction pattern showing hexagonal periodicity that is indexed according to the crystal structure shown in (a). (f) Simulated electron diffraction pattern.}
\end{figure}

\newpage

\begin{figure}
\includegraphics[width=1.0\linewidth]{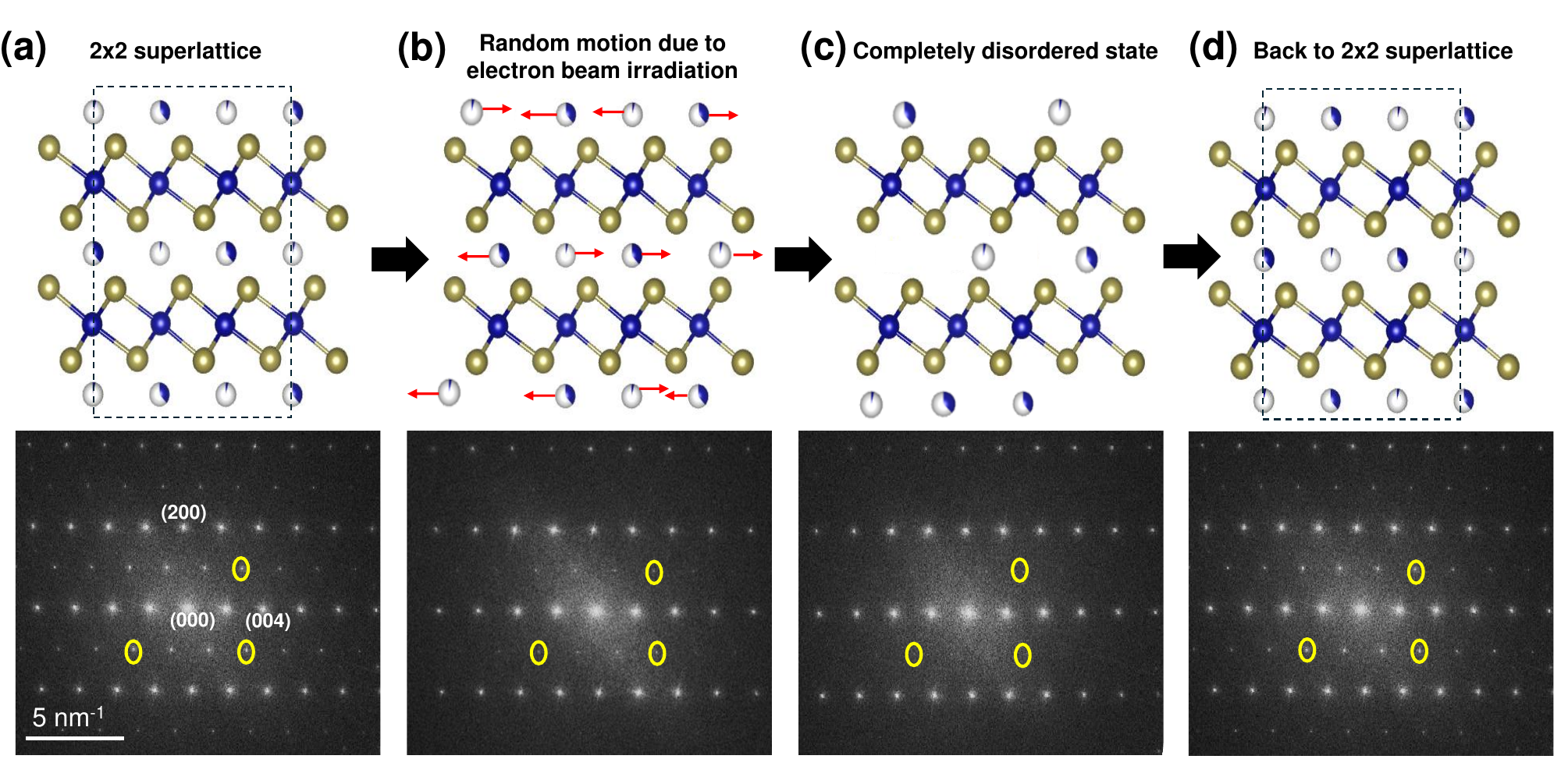}
\caption{(a) Side view of the crystal structure of Cr$_{4.89}$Te$_8$ as determined by XRD (top) and fast Fourier transform (FFT) of the TEM image taken along the (010) direction (bottom). The spots highlighted by the yellow circles in the FFT measurement correspond to the 2$\times$2 superlattice periodicity of Cr intercalants. (b,c) Same as (a) but taken from samples after 2 and 10 minutes under electron beam irradiation in TEM. (d) The TEM image was taken from the sample after the electron beam was stopped for several minutes.} 
\end{figure}

\newpage
\begin{figure}
\includegraphics[width=1.0\linewidth]{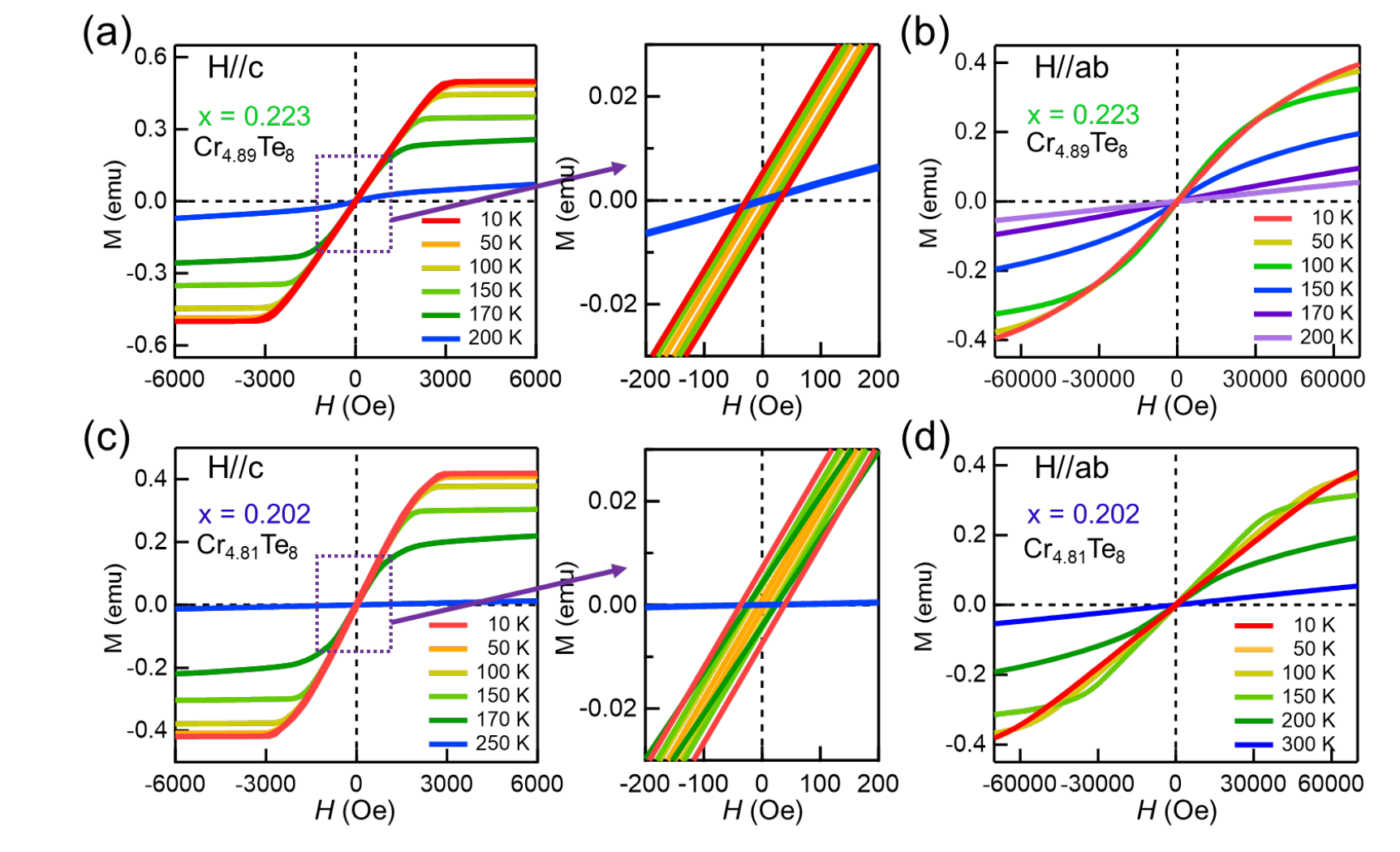}
\caption{(a) Magnetization of Cr$_{4.89}$Te$_8$ as a function of the OOP magnetic field strength at different temperatures. The zoom-in highlights the small coercive field in the hysteresis loop. (b) Same as (a) but measured with IP magnetic fields. (c,d) Same as (a,b) but for Cr$_{4.81}$Te$_8$.}
\end{figure}

\newpage
\begin{figure}
\includegraphics[width=1.0\linewidth]{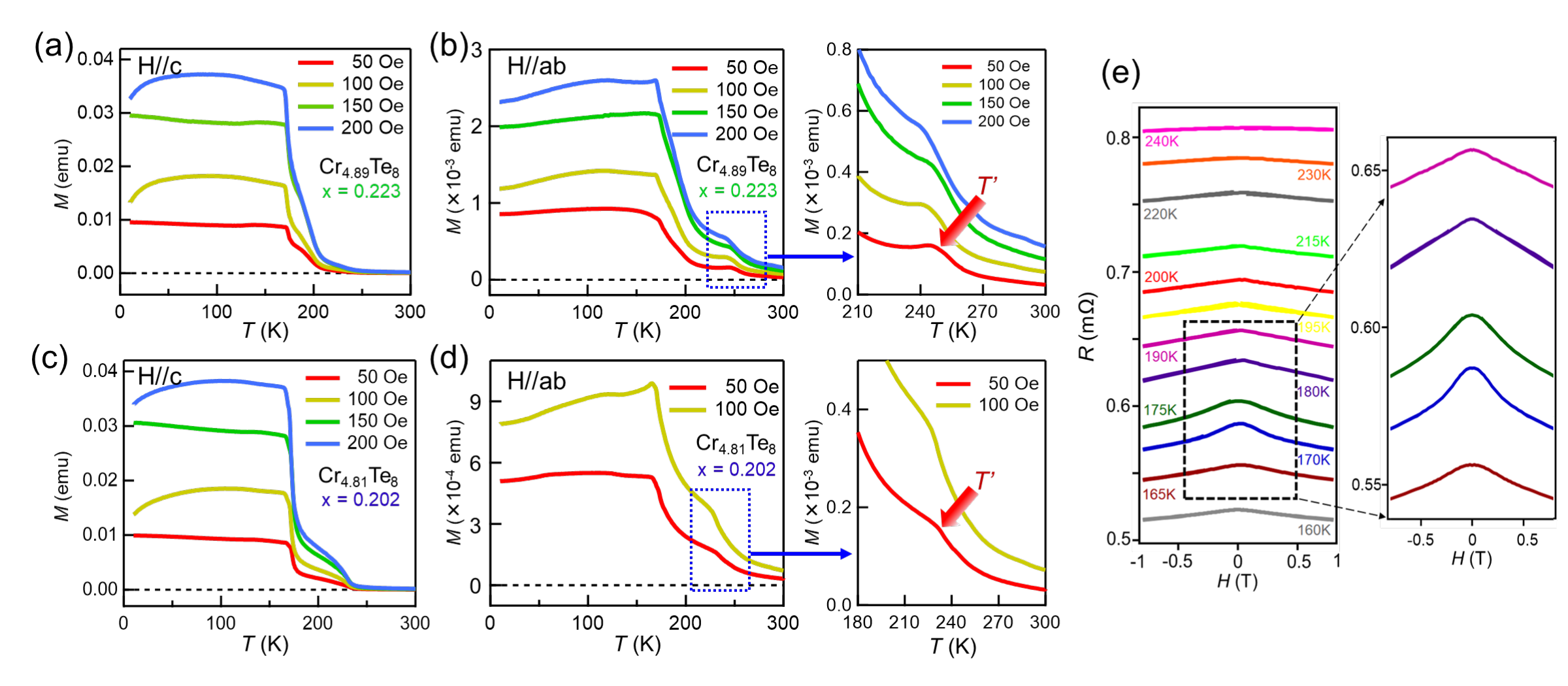}
\caption{ (a) Magnetization of Cr$_{4.89}$Te$_8$ as a function of temperature measured with different OOP magnetic fields. (b) Magnetization of Cr$_{4.89}$Te$_8$ as a function of temperature measured with different IP magnetic fields. The zoom-in highlights the bump feature in the $M$-$T$ curve, which indicates the emergent IP spin fluctuation in Cr$_{4.89}$Te$_8$. (c,b) Same as (a,b) but for Cr$_{4.81}$Te$_8$.  (e) Measured longitudinal resistance of Cr$_{4.81}$Te$_8$ as a function of the strength of the OOP magnetic field measured at different temperatures.
}%
\end{figure}

\begin{table}[h]
\includegraphics[width=0.95\linewidth]{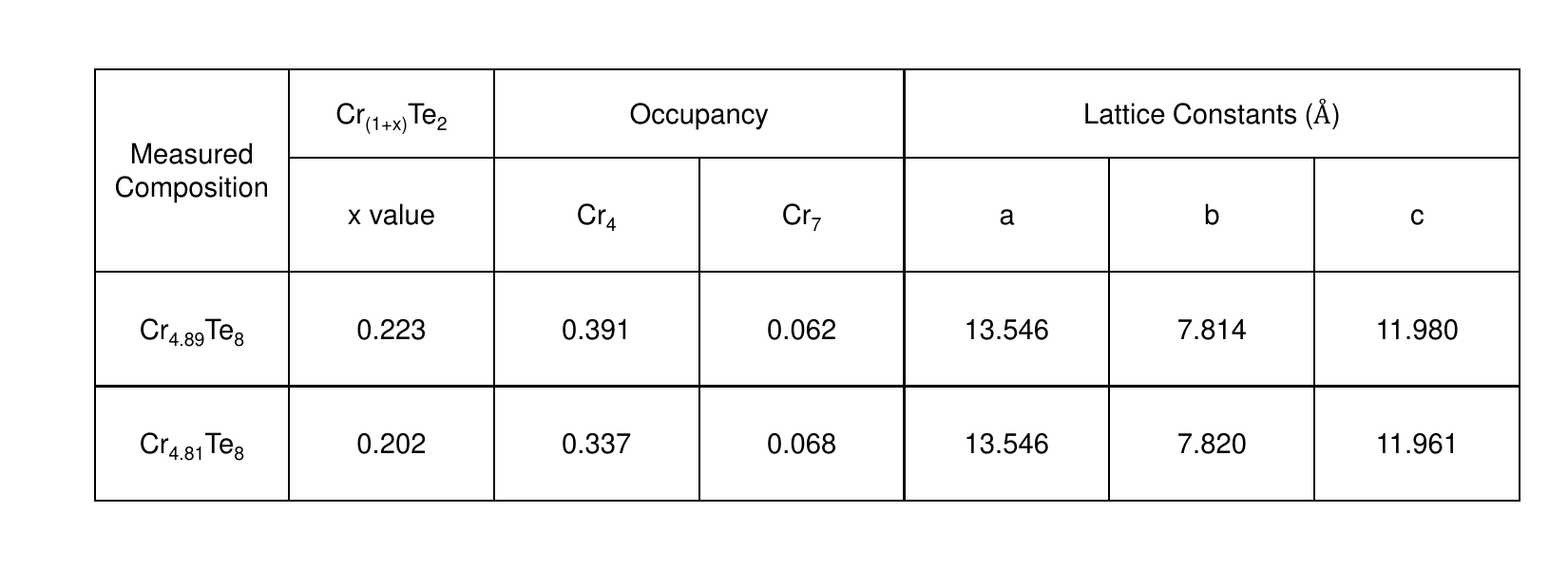}
\caption{Chemical composition and lattice constants of the Cr$_{1+x}$Te$_2$ samples measured by XRD.}%
\end{table}

\end{document}